\documentclass[sigconf, prologue, table, xcdraw]{acmart}
\usepackage{booktabs}
\usepackage{caption}
\usepackage{subcaption}
\usepackage{multirow}
\usepackage[utf8]{inputenc}
\usepackage{listings}
\usepackage{mathtools}
\usepackage{enumitem}
\usepackage{float}
\setcopyright{none}
\usepackage{csquotes}

\usepackage{hyphenat}
\usepackage[]{algorithm2e}
\usepackage[ flushmargin]{footmisc}

\newcommand{\hide}[1]{}
 \def\@copyrightspace{\relax}
\newcommand{\xhdr}[1]{\vspace{1.7mm}\noindent{{\bf #1.}}}

\newcommand{\Secref}[1]{Sec.~\ref{#1}}

\newcommand{\Tabref}[1]{Table~\ref{#1}}
\newcommand{\Figref}[1]{Fig.~\ref{#1}}

\newcommand{\ie}{{i.e.}\xspace}
\newcommand{\eg}{{e.g.}\xspace}

\newcommand{\etal}{{et al.}\xspace}
\newcommand{\vs}{{vs.}\xspace}

\copyrightyear{2019}
\acmYear{2019} 
\setcopyright{iw3c2w3}
\acmConference[WWW '19]{Proceedings of the 2019 World Wide Web Conference}{May 13--17, 2019}{San Francisco, CA, USA}
\acmBooktitle{Proceedings of the 2019 World Wide Web Conference (WWW '19), May 13--17, 2019, San Francisco, CA, USA}
\acmPrice{}
\acmDOI{10.1145/3308558.3313531}
\acmISBN{978-1-4503-6674-8/19/05}

\begin{document}

\title{Message Distortion in Information Cascades}

\author{Manoel Horta Ribeiro}
\authornote{\vspace{-1mm} Work done during an internship at EPFL. Code/data: \texttt{github.com/epfl-dlab/mdic}}
\affiliation{
\institution{UFMG}
}
\email{manoelribeiro@dcc.ufmg.br}

\author{Kristina Gligori\'c}
\affiliation{
\institution{EPFL}
}
\email{kristina.gligoric@epfl.ch}

\author{Robert West}
\affiliation{
\institution{EPFL}
}
\email{robert.west@epfl.ch}

\renewcommand{\shortauthors}{
M. Horta Ribeiro, K. Gligori\'c, and R. West
}

\begin{abstract}
Information diffusion is usually modeled as a process in which immutable pieces of information propagate over a network.
In reality, however, messages are not immutable, but may be morphed with every step, potentially entailing large cumulative distortions.
This process may lead to misinformation even in the absence of malevolent actors, and understanding it is crucial for modeling and improving online information systems.
Here, we perform a controlled, crowdsourced experiment in which we simulate the propagation of information from medical research papers. Starting from the original abstracts, crowd workers iteratively shorten previously produced summaries to increasingly smaller lengths.
We also collect control summaries where the original abstract is compressed directly to the final target length.
Comparing cascades to controls allows us to separate the effect of the length constraint from that of accumulated distortion.
Via careful manual coding, we annotate lexical and semantic units in the medical abstracts and track them along cascades. We find that iterative summarization has a negative impact due to the accumulation of error, but that high-quality intermediate summaries result in less distorted messages than in the control case.
Different types of information behave differently; in particular, the conclusion of a medical abstract (\ie, its key message) is distorted most.
Finally, we compare abstractive with extractive summaries, finding that the latter are less prone to semantic distortion.
Overall, this work is a first step in studying information cascades without the assumption that disseminated content is immutable, with implications on our understanding of the role of word-of-mouth effects on the misreporting of science.
\end{abstract}

%
%



\maketitle

\section{Introduction}
\authornote{Work done during an internship at EPFL. Code and data for this paper are available at \texttt{https://github.com/epfl-dlab/mdic}}
\begin{figure}[t]
\centering
\includegraphics[width=0.64\linewidth]{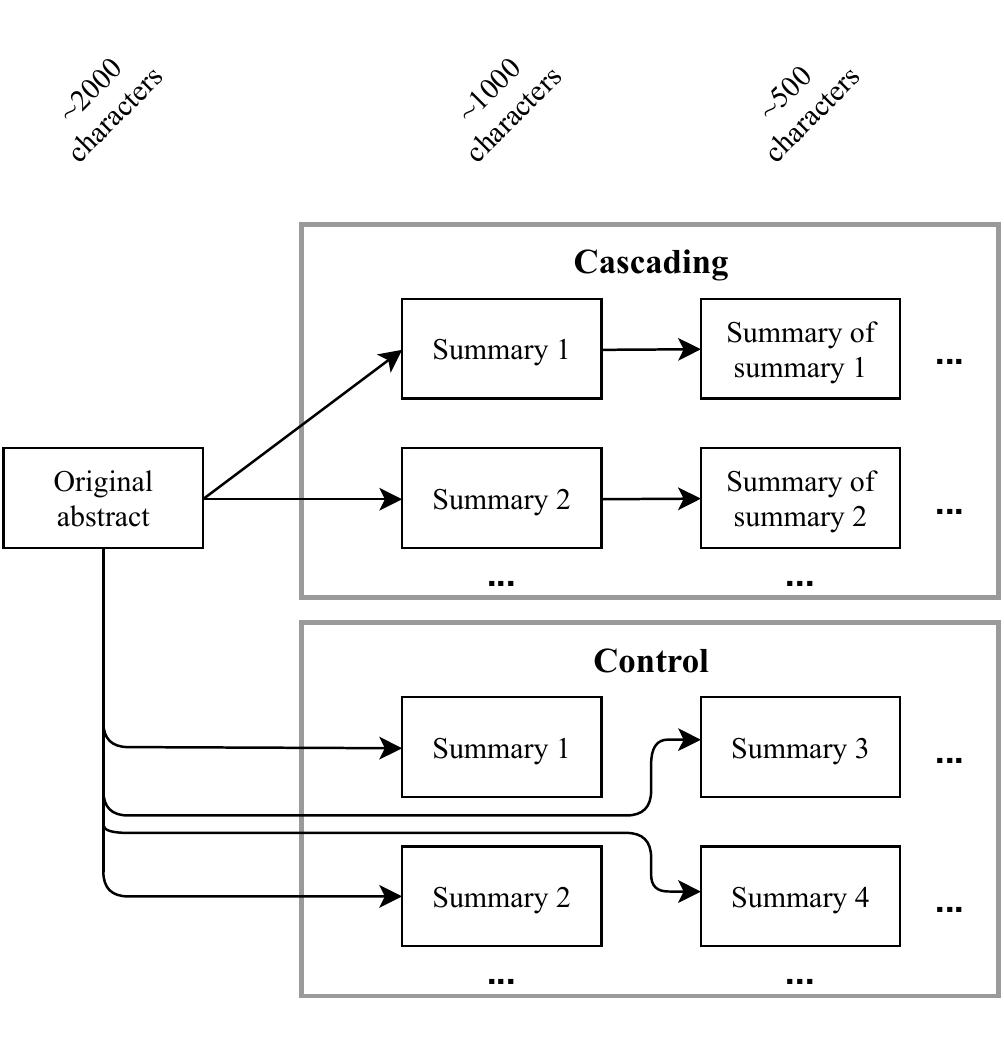}
\caption{Schema of the crowdsourced experiment for simulating information cascades.
In the cascading setting, workers summarize texts iteratively, reducing the number of characters hop by hop.
In the control setting, workers always summarize the original text for all target lengths.}
\label{fig:experiment-setting}
\end{figure}

The spread of information, online and offline, is a noisy process.
As a message is passed on from person to person, or from platform to platform, errors creep in, and facts are distorted, oftentimes to an extent such that, after a few hops of propagation, downstream messages may be entirely different from---or even contradict---the original message.
This way, valuable information may turn into harmful misinformation, even without purposeful interference.

Misconceptions and sensationalism add their part to compound the problem, as frequently observed in the context of health\hyp related topics such as dieting and vaccination. 
For example, in 2006, the first results of the \textit{Women's Health Initiative Dietary Intervention} trial were published.
The trial found little impact of diets lower in fat and higher in fruits, vegetables, and grains in the incidence of certain diseases in women between 50 and 79 years old~\citep{howard_low-fat_2006}.
Shortly after its publication, a sequence of press releases and news stories increasingly distorted the nuanced and cautious findings of the study, overlooking methodological caveats and benefits found~\citep{lissner_womens_2006}.
Throughout the diffusion process, news overwhelmingly reported that food and nutrition had little to do with health and disease~\citep{brody_more_2015}.

This anecdote portrays how information may be distorted as it propagates over the news and social media.
These distortion processes are overlooked in the existing literature on information diffusion~\citep{guille_information_2013}, which treats information disseminated through a network as consisting of immutable pieces of content (\eg,
\textit{memes} ~\citep{leskovec_meme-tracking_2009}
or topics~\citep{cataldi_emerging_2010}). 
Previous research in communication studies~\citep{hall_encoding/decoding_2009, entman_framing:_1993} indicates, however, that the way information is altered, interpreted, or framed along its diffusion may have a significant impact.

Two orthogonal factors are at play during message propagation:

\begin{enumerate}
    \item \textbf{Word of mouth:} Information commonly spreads in a cascading fashion, from person to person, or from platform to platform, rather than directly from the original source to every person or platform.
    \item \textbf{Summarization:} When an original message is passed on, it is frequently compressed, focusing on the essence while omitting unnecessary details.
\end{enumerate}

Both factors can introduce errors.
First, word-of-mouth propagation usually takes place on noisy channels (unless messages are forwarded unmodified, \eg, via retweets), and when an error occurs, it is passed on via what we term the \textbf{telephone effect,} named after the \textit{telephone game,} in which \textit{``players form a line, and the first player comes up with a message and whispers it to the ear of the second person in the line. The second player repeats the message to the third player, and so on. When the last player is reached, they announce the message they heard to the entire group. The first person then compares the original message with the final version. Although the objective is to pass around the message without it becoming garbled along the way, this usually ends up happening''}~\citep{noauthor_chinese_2018}.
Second, summarization can be seen as lossy compression and thus induces an additional loss of information, which we term the \textbf{summary effect.}

Consider, \eg, this three-hop cascade: a ten-page medical research paper is promoted in a one-page university press release, which is picked up by a half-page newspaper article, which is finally mentioned in a tweet with 280 characters.
It is clear that the tweet at the end of the cascade will be different from the original research paper.
What is less clear is whether the difference stems from the fact that the message was passed on three times (the telephone effect), or from the fact that ten pages were compressed to 280 characters (the summary effect).
Disentangling the telephone and summary effects is difficult when working with observationally collected information cascades as studied in prior work, \eg, URLs spreading via tweets~\citep{rodrigues_word--mouth_2011} or quotations spreading via news articles and blog posts~\citep{leskovec_meme-tracking_2009}.
Moreover, assembling an appropriate dataset in the first place is challenging, too.
For instance, in the case of the aforementioned \textit{Women's Health} trial, it is hard even to identify the structure of cascades, \ie, the graph that specifies from which other node each node received the message.
If an article misreported the findings, it is unclear if the author read the original scientific report and misunderstood it, or if they read other articles that had already introduced the error.
Also, the level of coverage may vary widely: a special feature on women's health in a tabloid may briefly mention the trial, whereas an editorial in \textit{Science} may be dedicated to it entirely. Meaningfully comparing such different formats is hard.

\xhdr{Present work: an experimental framework for studying message distortion in information cascades}
These challenges associated with observational settings motivated us to adopt an experimental approach.
In our experimental design, inspired by the telephone game, we aim to track the distortion of messages as they propagate hop by hop.
Starting from an original message, we simulate an information cascade by asking a crowd worker to shorten the original message to a prescribed target length while preserving the essential information.
The resulting summary is then passed on to another worker, who is asked to condense it to an even shorter target length, and so forth.
This way, we obtain chains of \textbf{cascading summaries.} Along the chains, the original message is distorted by the telephone and summary effects.
To tease the two effects apart, we also collect \textbf{control summaries,} of the same lengths as the cascading summaries, but produced by directly summarizing the original message to the respective target length, without any intermediate summaries.
This setup is depicted in \Figref{fig:experiment-setting}.
Cascading summaries are subject to both the telephone and the summary effects, whereas control summaries are only subject to the summary effect, so the difference in error rates in cascading \vs\ control summaries can be ascribed to the telephone effect.

\begin{table*}[t]
\scriptsize
\caption{Papers used and associated topics: vaccination (VA), breast cancer (BC), cardiovascular diseases (CD), nutrition (NU).}
\label{tab:papers}
\begin{tabular}{p{7.5cm}p{0.5cm}|p{7.5cm}p{0.5cm}}
\toprule
\textbf{Paper} & \textbf{Topic} & \textbf{Paper} & \textbf{Topic} \\ \midrule
A population-based study of measles, mumps, and rubella vaccination and autism~\citep{madsen_population-based_2002} &
VA
& Effect of rosiglitazone on the risk of myocardial infarction and death from cardiovascular causes~\citep{nissen_effect_2007}
&  CD
\\ \midrule
Response to a monovalent 2009 influenza A (H1N1) vaccine~\citep{greenberg_response_2009} &
VA
& Azithromycin and the risk of cardiovascular death~\citep{ray_azithromycin_2012}
&  CD
\\ \midrule
First results of phase 3 trial of RTS, S/AS01 malaria vaccine in African children~\citep{rts_first_2011} &
VA
& Global sodium consumption and death from cardiovascular causes~\citep{mozaffarian_global_2014} &
CD \\ \midrule
Waning protection after fifth dose of acellular pertussis vaccine in children~\citep{klein_waning_2012}      & 
VA
&
Effect of sibutramine on cardiovascular outcomes in overweight and obese subjects~\citep{james_effect_2010} &
CD
\\ \midrule
Adjuvant exemestane with ovarian suppression in premenopausal breast cancer~\citep{pagani_adjuvant_2014} &       
BC
& Changes in diet and lifestyle and long-term weight gain in women and men \citep{mozaffarian_changes_2011}
& NU \\ \midrule
Effect of screening mammography on breast-cancer mortality in Norway~\citep{kalager_effect_2010} &
BC
& Comparison of weight-loss diets with different compositions of fat, protein, and carbohydrates~\citep{sacks_comparison_2009}  
& NU \\ \midrule
Exemestane for breast-cancer prevention in postmenopausal women~\citep{goss_exemestane_2011} & 
BC
& Primary prevention of cardiovascular disease with a Mediterranean diet~\citep{estruch_primary_2013}
& NU \\ \midrule
Effect of three decades of screening mammography on breast-cancer incidence~\citep{bleyer_effect_2012} &
BC
&  Association of coffee drinking with total and cause-specific mortality~\citep{freedman_association_2012}     
& NU \\ \bottomrule
\end{tabular}
\end{table*}

This experimental design allows us to address the following \textbf{research questions:}

\begin{enumerate}
    \item[\textbf{RQ1:}] \textbf{Measuring the telephone effect.} What part of information distortion is due to the cascading of messages (telephone effect), rather than to length restrictions (summary effect)?
    \item[\textbf{RQ2:}] \textbf{Information persistence.} Given that a piece of information has already survived $k$ summarization steps, how likely is it to survive one more? What factors impact its survival?
    \item[\textbf{RQ3:}] \textbf{Extractive \vs\ abstractive summarization.} Broadly, there are two ways of summarizing text,
    (1)~by subselecting keyphrases (\textit{extractive}),
    (2)~by paraphrasing essential information (\textit{abstractive}). 
    How effective are these strategies in mitigating distortion introduced by the telephone effect?
\end{enumerate}

\vspace{-1mm}

\xhdr{Application to medical information}
Motivated by the importance of medical information, we apply the above framework to a scenario where original messages are abstracts of papers published in the \textit{New England Journal of Medicine} (NEJM).
Automatically identifying the key facts contained in a medical abstract is an open challenge \cite{nye_corpus_2018}, and so is determining the presence \vs\ absence of those facts in subsequent summaries.
Hence, in order to reliably address the above research questions, we manually annotate lexical and semantic units in the abstracts and track them along cascades.

We find that, overall, cascading summarization has a detrimental effect due
to the accumulation of error.
High-quality intermediate summaries, however, can have a positive effect, by isolating the essential information and discarding noise, entailing less distorted subsequent messages than in the control case.
Different types of information behave differently; in particular, the
conclusion of a medical abstract---the most critical information---is
distorted most: the conclusion is correctly represented in cascading summaries by about 25 percentage points less, compared to control summaries, for a fixed target length.
Moreover, we find that the prior knowledge of the crowd workers  impact the persistence of information in the control setting, but not in the cascading setting.
Finally, comparing extractive with abstractive summaries, we find that extractive summaries (where more keyphrases are preserved) are less prone to semantic distortion.

\xhdr{Implications}
Beyond the special case of medical information, our work has general implications for the study of information cascades in a setting where information is not immutable and atomic but subject to distortion and omission.
In this context, modeling the distortion of content through its diffusion in a network can help us understand the nature of viral content and the processes through which biased or erroneous information arises from a correct source.
Furthermore, these insights could be used to purposefully create content that is less prone to distortion as it is diffused.

\label{sec:intro}

\section{Research design}
We describe our experimental design for collecting cascading and control summaries (\Secref{sec:collecting_cascades}), the dataset of cascades of medical information (\Secref{sec:dataset_medical}), and our method for extracting keyphrases and facts from abstracts and tracking them along cascades (\Secref{sec:annotating_tracking}).

\subsection{Collecting information cascades}
\label{sec:collecting_cascades}

We perform a controlled experiment to simulate an information diffusion process through text summarization tasks. Starting from an original text of length $l_0$, we leverage crowd workers to shorten it to a sequence of prescribed target lengths $l_1 > l_2 > \dots > l_n$. More specifically, workers were given the following instructions:

\vspace{1mm}
\noindent\fbox{%
    \parbox{0.96\linewidth}{%
    \scriptsize
    You will be given a short text ($l_k$ characters) with medicine-related information. Your tasks are these:
    \begin{enumerate}
        \item     Read the text carefully.
        \item     Write a summary of the text. Your summary should
    \begin{enumerate}
        \item     convey the most important information in the text, as if you are trying to inform another person about what you just read;
        \item     contain between $l_{k+1} - \Delta_{k+1}$ and $l_{k+1} + \Delta_{k+1}$ characters.
    \end{enumerate}
    \end{enumerate}
    We expect high-quality summaries and will manually inspect some of them. Copy-pasting is disabled.
    }%
}
\vspace{1mm}

Cascading and control summaries differ with respect to the input given to workers. In cascading summaries, a crowd worker whose task is to produce a summary of length $l_k$ is given as input a summary of length $l_{k-1}$ that was previously produced by another crowd worker. Only for target length $l_1$ is the original text used as input. In control summaries, on the contrary, the input is always the original text, regardless of the target length.

For each original text, target length, and condition, we collect multiple summaries. Cascading summaries form a set of chains rooted in the original abstracts, whereas control summaries are flat broadcast trees. This is depicted in \Figref{fig:experiment-setting}.

\begin{figure*}[ht]
\centering
\includegraphics[width=0.85\linewidth]{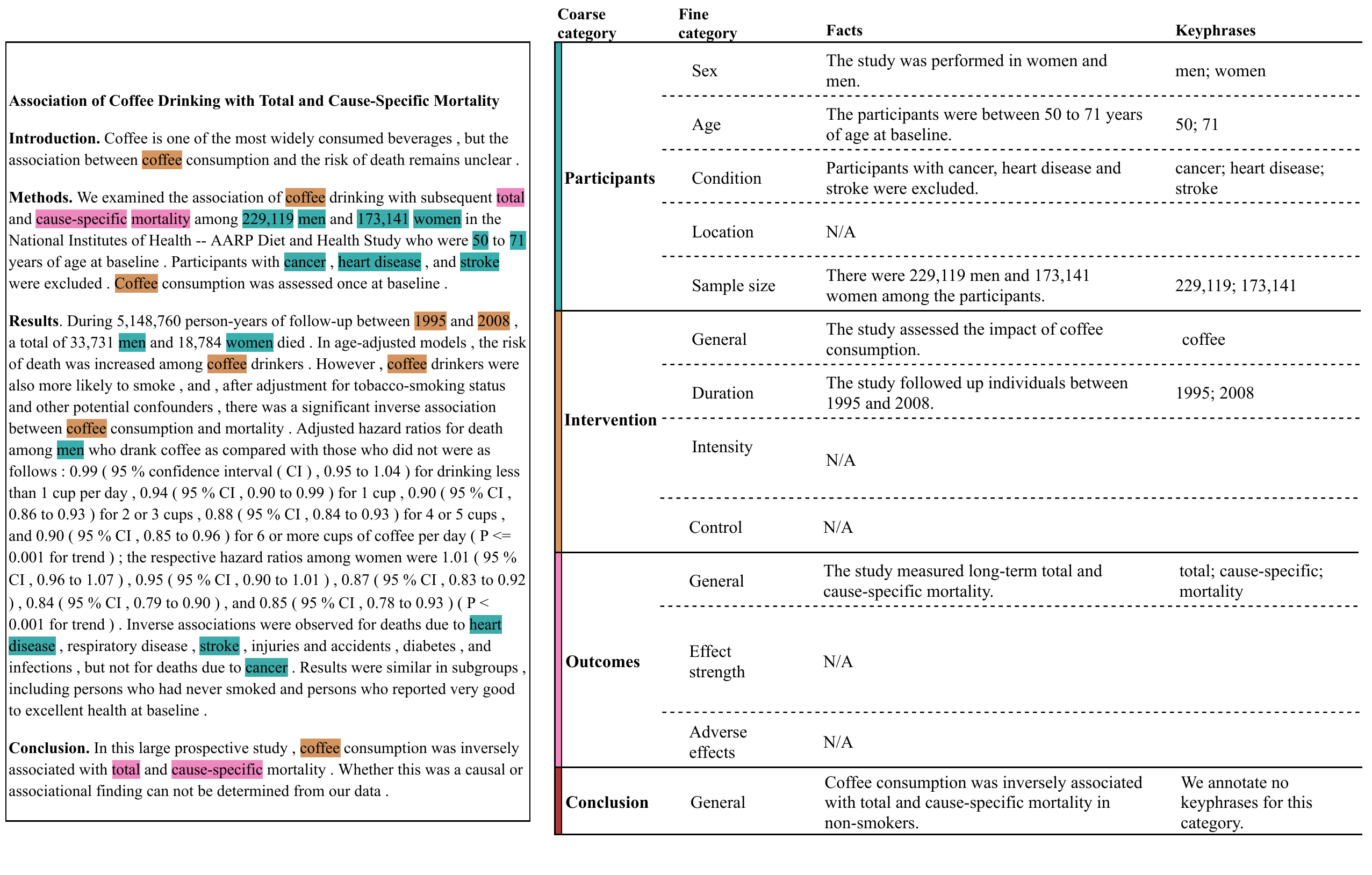}
\caption{
\textit{Left:} one of the 16 abstracts used in this study.
\textit{Right:} the categories employed, as well as facts and keyphrases associated with the abstract. Keyphrases are highlighted in the abstract.}
\label{fig:table_doggos_tokens}
\end{figure*}

Cascading summaries are affected by both the telephone and the summary effect (as defined in \Secref{sec:intro}), whereas control summaries are affected only by the summary effect, such that comparing the two cases allows us to quantify the telephone effect.
One might argue that the telephone effect could be isolated in a more straightforward fashion by using a constant target length throughout, asking workers to simply rephrase the input, and thus eliminating the summary effect altogether.
This, however, would allow for trivial solutions on behalf of crowd workers: nothing would keep them from simply copying the input, such that by eliminating the summary effect, the telephone effect would also vanish.
This shortcoming could be addressed using a different research design, \eg, by showing workers the input text for a certain amount of time, then hiding it and asking them to reproduce it from memory. This setup, however, is more complex to implement, requires workers to spend idle time before finishing the task, and runs the risk of cheating (\eg, even when disabling copy--paste, users might take a screenshot or a picture). 
Also, for longer original texts (such as research paper abstracts), summarization is a step users would naturally perform in real cascades. For all these reasons, we adopt the summarization-based paradigm. 
Due to budget constraints, we create cascades such that each node has one descendant, since otherwise, the cost of the experiment would increase exponentially.

\begin{figure*}[t]
\centering
\includegraphics[width=0.92\linewidth]{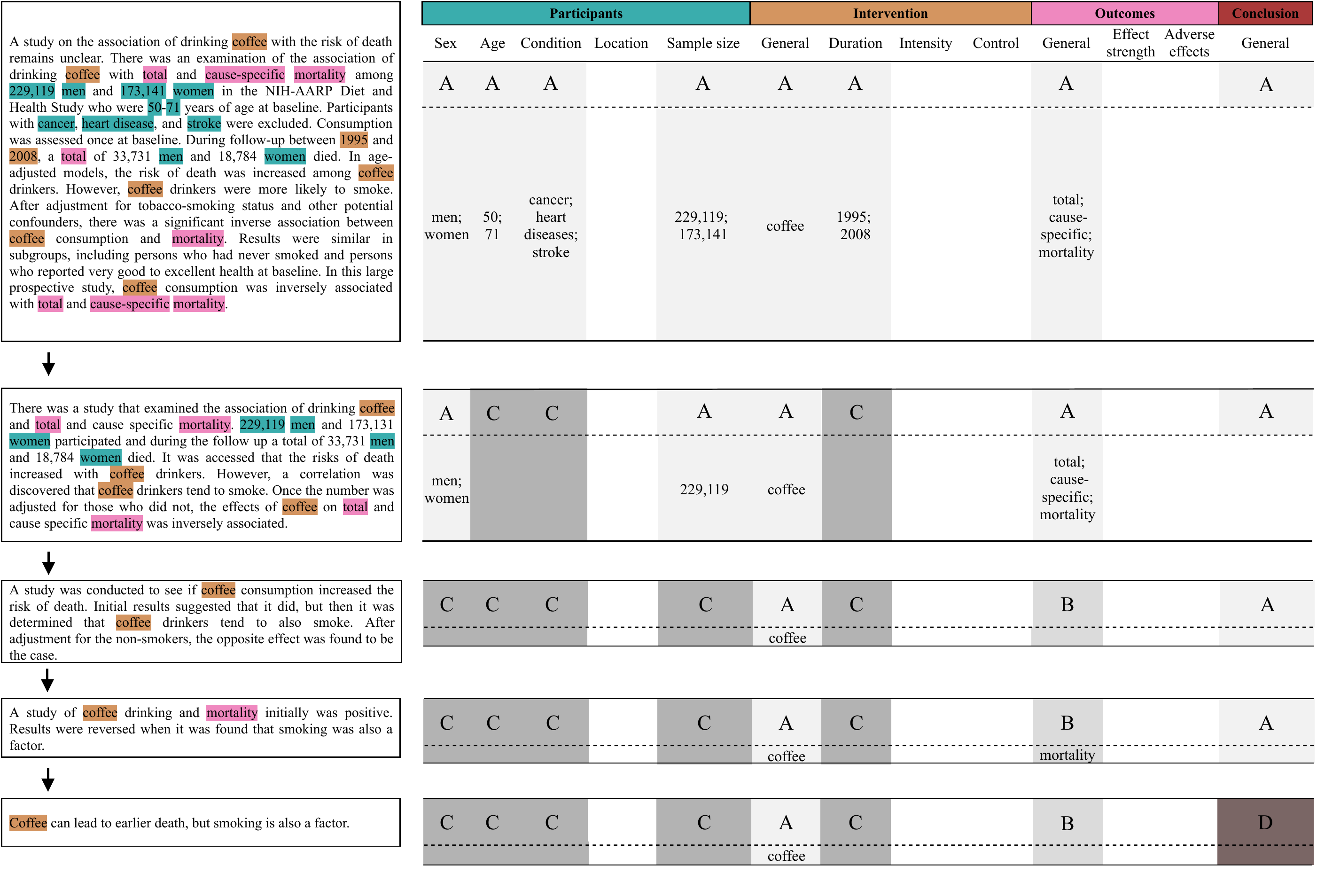}
\caption{
\textit{Left:} a real summarization cascade obtained from crowd workers.
\textit{Right:} the annotated facts and the matched keyphrases for each of the summaries. Fact values (\textit{A,} \textit{B,} \textit{C,} \textit{D}) are explained in \Tabref{tab:descriptions}.
}
\label{fig:example_cascades}
\end{figure*}

\subsection{Dataset: cascades of medical information}
\label{sec:dataset_medical}

In our concrete application of the above experimental design, we collect cascades whose root nodes consist of medical abstracts from the \textit{New England Journal of Medicine} (NEJM).
We select four research fields of public interest (vaccination, breast cancer, cardiovascular diseases, and nutrition), and choose 4 impactful papers per field, for a total of 16 abstracts, listed in \Tabref{tab:papers}. Analyzing the number of links to these papers in a large corpus of blog posts and news pieces, we select papers that were widely discussed online.

For each abstract, we generate $8$ chains of cascading summaries, so that for each target length, each abstract had $8$ summaries associated with it. In parallel, we collect another $8$ independent control summaries per target length, totaling $16 \times 8 \times 2=256$ summaries per target length.
Original abstracts were $l_0 \approx 2000$ characters long, and we consider five target lengths (which we also refer to as \textit{hops}):
$l_1=1000$, $l_2=500$, $l_3=250$, $l_4=125$, and $l_5=64$ characters. We allow slacks $\Delta_1 = 100$, $\Delta_2 = 50$, $\Delta_3 = 25$,  $\Delta_4 = 13$ and  $\Delta_5 = 9$, for each budget respectively.

For several reasons, we enforce that workers only summarize one text per hop per abstract. Firstly, we do not want a single worker to be involved in several summarization chains for a given hop, as this would imply that a single unskilled or malicious worker could jeopardize the quality of all the chains. Secondly, we do not want information to leak from one summary to another. As different chains are related to the same paper, it could be that workers understood the text better if they summarized multiple summaries originating from the same paper. Due to the latter reason, we also limit workers to do summaries related to a paper in different hops $36$ hours apart.

To assess participants' level of domain knowledge, we use a questionnaire for each topic.
For the topics \textit{Breast cancer} and \textit{Vaccination,} we use online quizzes approved by University of Rochester medical reviewers~\citep{arnold_immunization_nodate, adam_s_breast_nodate}.
For the topic \textit{Cardiovascular diseases}, we use the sections \emph{Epidemiology} and \emph{Risk factors} of Bergman \etal \citep{bergman_development_2011}. For the topic \textit{Nutrition,} we use the first section on expert nutrition advice of the UCL Nutrition Knowledge Questionnaire~\citep{kliemann_reliability_2016}.

\subsection{Annotating and tracking information}
\label{sec:annotating_tracking}

Studying information distortion requires quantifying the completeness and truthfulness of a summary: which facts from the original text are present in a given summary, and which facts have been omitted or even contradicted?

To address this challenge, we exploit the fact that medical abstracts are highly structured: key information can be attributed to a few main categories (\eg, \textit{Participants} or \textit{Intervention}), which may be further decomposed into multiple lower\hyp level ones~\cite{nye_corpus_2018}. For example, \textit{Participants} may be decomposed into subcategories like \textit{Age,} \textit{Sex,} and \textit{Condition}.
We develop two methods for tracking information related to these categories, which we explain in detail.

We annotate \textbf{keyphrases} (\eg, ``mortality'') in the abstracts with the categories they belong to (\eg, \textit{Outcomes}). The phrases are then matched in subsequent summaries. Tracking keyphrases is computationally simple, but runs the risk of low recall: a fact may be expressed in a summary, but in words different from those in the input text. In other words, semantic completeness does not require lexical completeness. Moreover, it may be the case that the keyphrase is present, but the actual information is lost, \eg, if the summary contradicts the source text but shares keyphrases with it. 

\begin{table}
\scriptsize
\caption{Values used for annotating facts.}
\label{tab:descriptions}
\begin{tabular}{p{0.9cm}p{6.7cm}}
\toprule
Fact value & Explanation \\ \midrule
 \textit{A}    & The fact is entirely captured in the text, omitting only insignificant details.             \\
 \textit{B}    & The essence of the fact is captured in the text, but a significant amount of detail was omitted.            \\
 \textit{C}    & The fact is not, or only insufficiently, captured in the text.            \\
 \textit{D}    & The fact contradicts the original text.           \\ \bottomrule
\end{tabular}
\end{table}

We also extract \textbf{facts,} short sentences stating the essential information a text conveys about a specific category. For example, for \textit{Participants\slash Sample size,} a fact may be \textit{``There were 229,119 men and 173,141 women among the participants''}. Facts of a given category may partially overlap with others; \eg, in the above example, it is mentioned that the study was performed with men and women, information which may also be present in the fact for the \textit{Participants\slash Sex} category. We present a full abstract with annotated keyphrases and facts in \Figref{fig:table_doggos_tokens}. On the right-hand side, we show the hierarchical categories for both keyphrases and facts, inspired by the categories proposed in Nye et al.~\cite{nye_corpus_2018}. Notice that \textit{Conclusion\slash General} is a category for which we have only facts but not keyphrases. This is the case as the conclusion often involves keyphrases from various categories; \eg, the statement \textit{``Coffee consumption was inversely associated with total and cause\hyp specific mortality in nonsmokers''} contains keyphrases about the intervention (\textit{coffee}) and the outcome (\textit{cause\hyp specific; mortality}).

Tracking facts along the cascade is more complex and ambiguous than tracking keyphrases. For each summary, we assign each fact a \textbf{fact value} (\textit{A,} \textit{B,} \textit{C,} or \textit{D}). The meaning of these values is defined in \Tabref{tab:descriptions}.
We attributed values to facts via crowdsourcing, instructing crowd workers as follows:

\vspace{1mm}
\noindent\fbox{%
    \parbox{0.96\linewidth}{%
    \scriptsize
    You will be given several statements (around 9) and a series of short texts (of different sizes) related to medicine. Your tasks are these:
    \begin{enumerate}
        \item     Read the statement and the texts carefully.
        \item     Judge how well the statement is captured by the different texts, choosing between \textit{A,} \textit{B,} \textit{C,} and \textit{D} (as explained in \Tabref{tab:descriptions}).
    \end{enumerate}
        The above will be repeated for several statements (the texts will be the same).
        Quality checks will be performed on the answers.
    }%
}
\vspace{1mm}

Attributing values to the facts is subjective: it is often not trivial to decide between $A$ and $B,$ or between $B$ and $D;$ \eg, the information that the study involved 1,067 participants can be distorted along the cascade by stating that the study involved a thousand participants. Is this fact partially preserved, as the order of magnitude is still conveyed, or does this contradict the initial fact, since 1,067 $\neq$ 1,000?

We take several steps to ensure annotation quality:
\textit{(i)} We add a qualification test where we explain the task thoroughly, gave examples, and assessed workers' ability to understand the subtleties of the task.
\textit{(ii)} We show texts of different sizes at once to ensure that workers could compare what a piece of information would look like in several lengths, making the annotation of a given fact consistent across several summaries.
\textit{(iii)} We introduce quality checks (facts that are wrong for all statements), allowing us to filter workers who failed often.
\textit{(iv)} Lastly, we assign three workers per fact and manually review all facts, giving emphasis to those where not all three workers had annotated a given fact with the exact same value ($33\%$ of all cases).

\xhdr{Example} We present a summary cascade on the left-hand side of \Figref{fig:example_cascades}.
The summaries refer to the abstract from \Figref{fig:table_doggos_tokens}. The findings are nuanced:
coffee is associated with an increased risk of death; yet, when accounting for confounding variables, such as smoking, the association is reversed.
In the cascade, as the summaries become shorter, this subtlety becomes increasingly difficult to grasp. In the last summary, we eventually even arrive at a contradiction.

\begin{figure*}[ht]
\centering
\includegraphics[width=\linewidth]{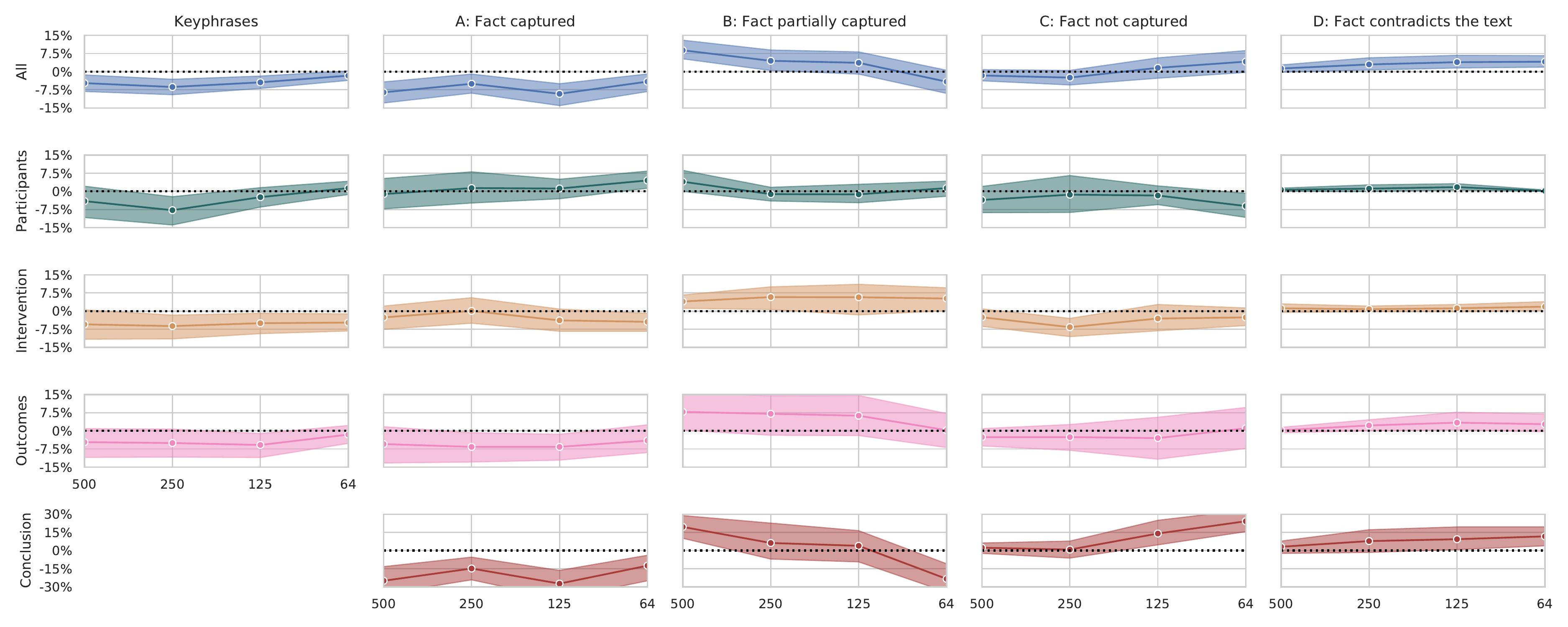}
\caption{Differences in percentage of preserved keyphrases (column~1) and fact values (columns 2--5) between cascading and control summaries for fact categories (rows) and hops ($x$-axes), with bootstrapped 95\% confidence intervals.
}
\label{fig:coarse-telephone}
\end{figure*}

Annotated keyphrases and extracted facts are shown on the right-hand side of \Figref{fig:example_cascades}. 
Firstly, notice that the sudden contradiction we just described is captured, as the label $D$ was assigned to the \textit{Conclusion\slash General} fact in the last summary.
Another interesting transition here is how the \textit{Outcomes\slash General} fact shifts from \textit{A} to \textit{B} in the third summary:
both previous summaries indicated that the summary measured total and cause-specific mortality, whereas the third one simply indicates that it measured the risk of death, which is less complete.

To see the limitations of tracking keyphrases and how they are overcome by tracking facts,
consider the second summary, which contains a typo: it says that the study was performed in 173,131 women instead of 173,141.
Also, the worker used the expression ``risk of death'' instead of ``mortality''.
These examples show that keyphrases can be difficult to track when there are typos or when expressions are rephrased.

\label{sec:experimental_design}

\section{Results}

\begin{figure*}[h]
\centering
\includegraphics[width=.95\linewidth]{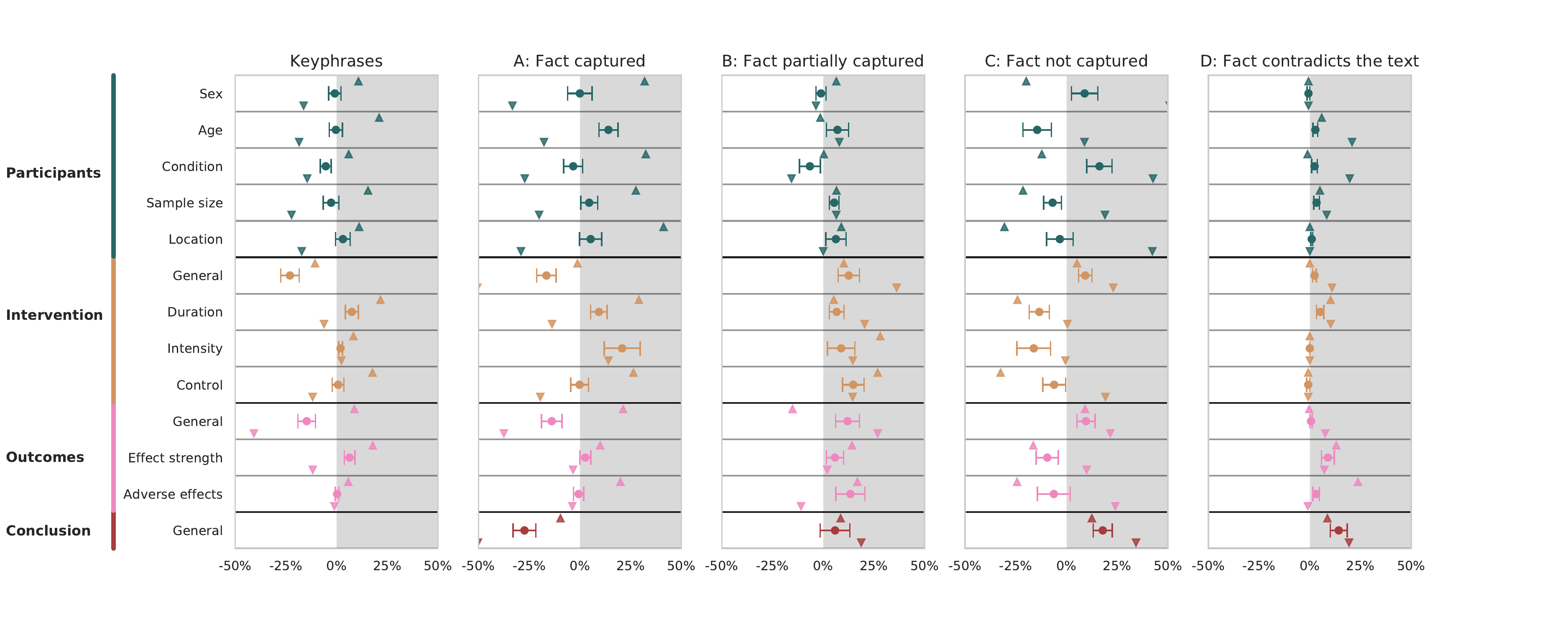}
\caption{Differences in the percentage of facts and keyphrases between cascading and control summaries for fine\hyp grained categories (averaged over all target lengths). On the positive side of the $x$-axis (in gray), the percentage of facts\slash keyphrases is higher in the cascading scenario. Each category is associated with three values (with bootstrapped 95\% CIs): The circle $\circ$ represents the actual difference between the cascading and control settings. The upward $\triangle$ (downward $\triangledown$) triangle represents the difference between the successor of the best (worst) cascading summary from the previous hop and control. 
}
\label{fig:tokens_fine}

\end{figure*}

\subsection{RQ1: Strength of the telephone effect}
\label{rq1}

In our first analysis, we compare the distortion of information in the cascading and control conditions.

\textbf{Differences across hops.} 
We calculate the percentage of keyphrases in the summaries of a given hop and the percentage of facts with a given value for all papers in the cascading and control settings.
We depict the difference between the two settings
in percentage points in \Figref{fig:coarse-telephone}. 
Values above $0\%$ mean that the percentage of keyphrases, or of facts of a given value, is higher in the cascading summaries. 
We observe several interesting patterns.
Cascading summaries preserve fewer keyphrases in all categories at all hops. This difference decreases in the last hops, perhaps because they require users to adopt more abstractive summarization strategies in both settings.
The percentage of facts labeled as \emph{A} is lower for cascading than for control summaries. This effect is particularly strong for the \textit{Outcomes} and \textit{Conclusion} categories. In the latter, the difference is greater than $15\%$.
The difference in the percentage of facts labeled as \textit{D} has an increasing trend. Here too, the effect is strongest for the \textit{Outcomes} and \textit{Conclusion} categories.
Lastly, facts labeled as either \textit{B} or \textit{C} do not present a clear trend for the \textit{Participants,} \textit{Intervention,} and \textit{Outcomes} categories. For the \textit{Conclusion} category, however, the difference in \textit{B}-valued facts decreases along the hops, while the percentage of \textit{C}-valued facts increases.
Overall, these differences suggest that the ``telephone effect'' impacts the quality of summaries significantly, particularly harming the essential facts---the outcomes and conclusion of a study.

\begin{figure}[h]
\centering
\includegraphics[width=\linewidth]{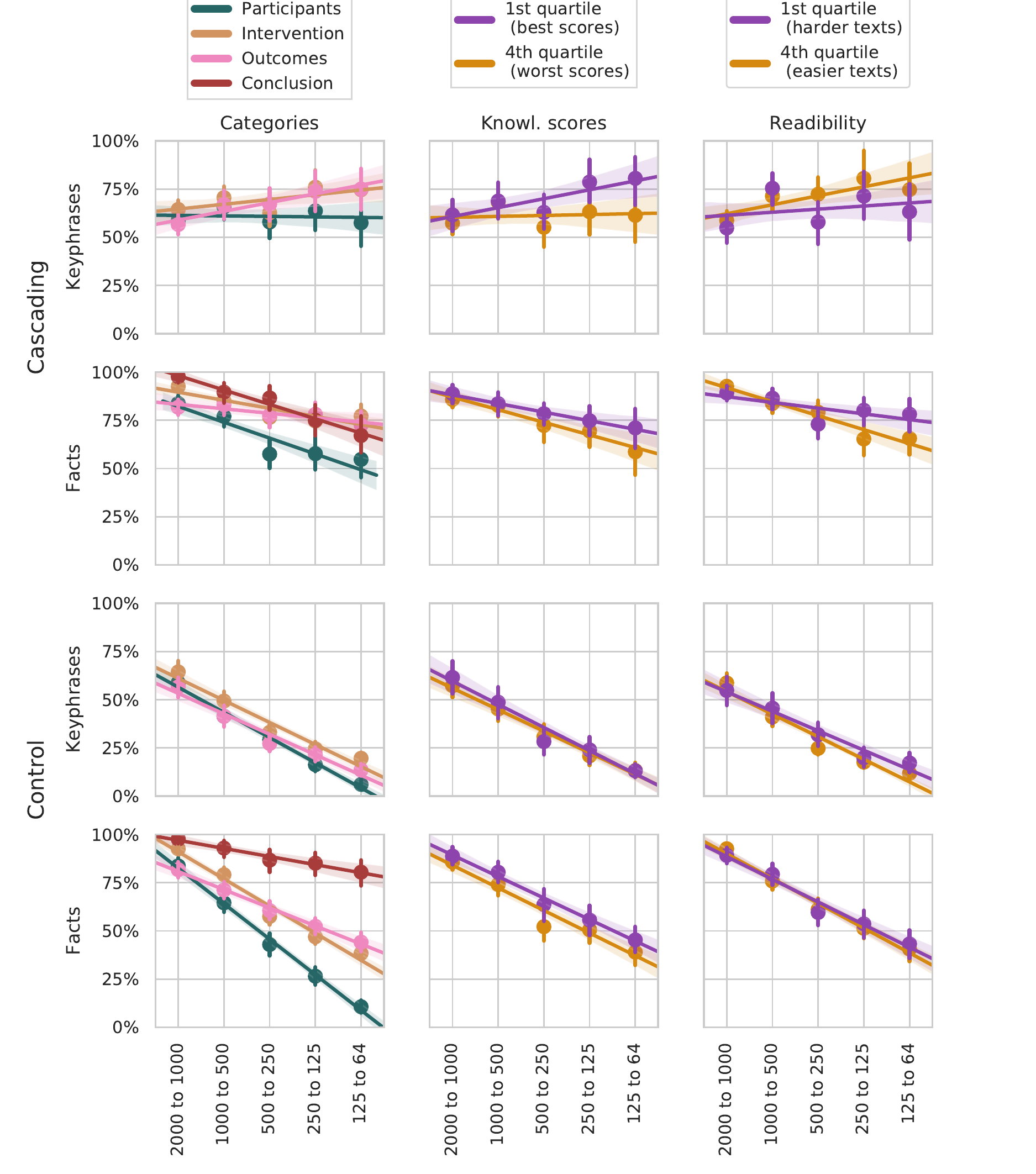}
\caption{
Probabilities of surviving one more hop for keyphrases\slash facts, alongside regression lines. Each row represents a combination of keyphrases\slash facts and a setting (cascading/control). In the first column, we compare the survival chance across different categories. In the other two, we compare it for different values of workers' questionnaire scores and of the readability index of the original abstract.
}
\label{fig:contintional}
\end{figure}

\textbf{Category distribution.} 
We also analyze how the distribution of categories differs for the distinct settings (\Tabref{tab:cat_distr}).
Consider the percentage of facts labeled as \emph{A} that belong to the \textit{Participants} and \textit{Conclusion} categories. For cascading summaries, the percentage of \textit{Participants} facts consistently stays above $29\%$, while this percentage drops significantly to $14\%$ for control summaries. For the \textit{Conclusion} category, we have the inverse: the percentage of $A$ statements in the \textit{Conclusion} category decreases in the cascading group and remains stable (above $10\%$) in the control. This suggests that the type of message that ends up getting spread differs between the cascading and control summaries: cascading summaries seem to ``remember'' less pertinent facts about participants, whereas control summaries preserve more crucial facts about study conclusions.

\begin{table}[H]
\small
\caption{Distribution over categories (\textit{\underline{P}articipants, \underline{I}n\-ter\-vention, \underline{O}utcomes, \underline{C}onclusion}) for fact values (\textit{A}, \textit{B}, \textit{C}, \textit{D}). Results for control setting in \textbf{bold}.}
\label{tab:cat_distr}
\begin{tabular}{llllll}
\hline
\multicolumn{2}{l}{\textbf{Tgt.\ len.}}           & \textbf{500}& \textbf{250}& \textbf{125}& \textbf{64}\\ \hline
\multicolumn{1}{|l|}{\cellcolor[HTML]{C0C0C0}}& \multicolumn{1}{c|}{\textbf{P}}
& \cellcolor[HTML]{EFEFEF}38\%/\textbf{35\%} & \cellcolor[HTML]{EFEFEF}36\%/\textbf{32\%} & \cellcolor[HTML]{EFEFEF}32\%/\textbf{25\%} & \multicolumn{1}{l|}{\cellcolor[HTML]{EFEFEF}29\%/\textbf{14\%}} \\ \cline{2-2}
\multicolumn{1}{|l|}{\cellcolor[HTML]{C0C0C0}}& \multicolumn{1}{c|}{\textbf{I}}
& 34\%/\textbf{33\%}   & 36\%/\textbf{34\%}   & 41\%/\textbf{37\%}   & \multicolumn{1}{l|}{50\%/\textbf{52\%}}   \\ \cline{2-2}
\multicolumn{1}{|l|}{\cellcolor[HTML]{C0C0C0}}& \multicolumn{1}{c|}{\textbf{O}}
& \cellcolor[HTML]{EFEFEF}17\%/\textbf{18\%} & \cellcolor[HTML]{EFEFEF}16\%/\textbf{18\%} & \cellcolor[HTML]{EFEFEF}18\%/\textbf{20\%} & \multicolumn{1}{l|}{\cellcolor[HTML]{EFEFEF}17\%/\textbf{21\%}} \\ \cline{2-2}
\multicolumn{1}{|l|}{\multirow{-4}{*}{\cellcolor[HTML]{C0C0C0}A}} & \multicolumn{1}{c|}{\textbf{C}} 
& 10\%/\textbf{14\%}   & 12\%/\textbf{16\%}   & 8\%/\textbf{18\%}   & \multicolumn{1}{l|}{4\%/\textbf{13\%}}   \\ \hline
\multicolumn{1}{|l|}{\cellcolor[HTML]{C0C0C0}}& \multicolumn{1}{c|}{\textbf{P}}
& \cellcolor[HTML]{EFEFEF}23\%/\textbf{23\%} & \cellcolor[HTML]{EFEFEF}17\%/\textbf{23\%} & \cellcolor[HTML]{EFEFEF}14\%/\textbf{19\%} & \multicolumn{1}{l|}{\cellcolor[HTML]{EFEFEF}13\%/\textbf{11\%}} \\ \cline{2-2}
\multicolumn{1}{|l|}{\cellcolor[HTML]{C0C0C0}}& \multicolumn{1}{c|}{\textbf{I}}
& 24\%/\textbf{26\%}  & 27\%/\textbf{21\%}   & 25\%/\textbf{20\%}   & \multicolumn{1}{l|}{24\%/\textbf{15\%}}   \\ \cline{2-2}
\multicolumn{1}{|l|}{\cellcolor[HTML]{C0C0C0}}& \multicolumn{1}{c|}{\textbf{O}}
& \cellcolor[HTML]{EFEFEF}35\%/\textbf{39\%} & \cellcolor[HTML]{EFEFEF}37\%/\textbf{37\%} & \cellcolor[HTML]{EFEFEF}38\%/\textbf{37\%} & \multicolumn{1}{l|}{\cellcolor[HTML]{EFEFEF}39\%/\textbf{36\%}} \\ \cline{2-2}
\multicolumn{1}{|l|}{\multirow{-4}{*}{\cellcolor[HTML]{C0C0C0}B}} & \multicolumn{1}{c|}{\textbf{C}} 
& 18\%/\textbf{12\%}   & 19\%/\textbf{19\%}   & 23\%/\textbf{25\%}   & \multicolumn{1}{l|}{24\%/\textbf{38\%}}   \\ \hline
\multicolumn{1}{|l|}{\cellcolor[HTML]{C0C0C0}}& \multicolumn{1}{c|}{\textbf{P}}
& \cellcolor[HTML]{EFEFEF}46\%/\textbf{47\%} & \cellcolor[HTML]{EFEFEF}48\%/\textbf{46\%} & \cellcolor[HTML]{EFEFEF}47\%/\textbf{47\%} & \multicolumn{1}{l|}{\cellcolor[HTML]{EFEFEF}45\%/\textbf{48\%}} \\ \cline{2-2}
\multicolumn{1}{|l|}{\cellcolor[HTML]{C0C0C0}}& \multicolumn{1}{c|}{\textbf{I}}
& 24\%/\textbf{24\%}  & 27\%/\textbf{29\%}   & 28\%/\textbf{29\%}   & \multicolumn{1}{l|}{28\%/\textbf{29\%}}   \\ \cline{2-2}
\multicolumn{1}{|l|}{\cellcolor[HTML]{C0C0C0}}& \multicolumn{1}{c|}{\textbf{O}}
& \cellcolor[HTML]{EFEFEF}27\%/\textbf{28\%} & \cellcolor[HTML]{EFEFEF}23\%/\textbf{23\%} & \cellcolor[HTML]{EFEFEF}20\%/\textbf{21\%} & \multicolumn{1}{l|}{\cellcolor[HTML]{EFEFEF}21\%/\textbf{21\%}} \\ \cline{2-2}
\multicolumn{1}{|l|}{\multirow{-4}{*}{\cellcolor[HTML]{C0C0C0}C}} & \multicolumn{1}{c|}{\textbf{C}} 
& 2\%/\textbf{1\%}   & 3\%/\textbf{2\%}   & 5\%/\textbf{2\%}   & \multicolumn{1}{l|}{6\%/\textbf{2\%}}   \\ \hline
\multicolumn{1}{|l|}{\cellcolor[HTML]{C0C0C0}}& \multicolumn{1}{c|}{\textbf{P}}
& \cellcolor[HTML]{EFEFEF}22\%/\textbf{20\%} & \cellcolor[HTML]{EFEFEF}21\%/\textbf{20\%} & \cellcolor[HTML]{EFEFEF}18\%/\textbf{0\%} & \multicolumn{1}{l|}{\cellcolor[HTML]{EFEFEF}2\%/\textbf{0\%}} \\ \cline{2-2}
\multicolumn{1}{|l|}{\cellcolor[HTML]{C0C0C0}}& \multicolumn{1}{c|}{\textbf{I}}
& 26\%/\textbf{20\%}  & 15\%/\textbf{15\%}   & 13\%/\textbf{17\%}   & \multicolumn{1}{l|}{15\%/\textbf{0\%}}   \\ \cline{2-2}
\multicolumn{1}{|l|}{\cellcolor[HTML]{C0C0C0}}& \multicolumn{1}{c|}{\textbf{O}}
& \cellcolor[HTML]{EFEFEF}19\%/\textbf{27\%} & \cellcolor[HTML]{EFEFEF}24\%/\textbf{20\%} & \cellcolor[HTML]{EFEFEF}24\%/\textbf{0\%} & \multicolumn{1}{l|}{\cellcolor[HTML]{EFEFEF}24\%/\textbf{18\%}} \\ \cline{2-2}
\multicolumn{1}{|l|}{\multirow{-4}{*}{\cellcolor[HTML]{C0C0C0}D}} & \multicolumn{1}{c|}{\textbf{C}} 
& 33\%/\textbf{33\%}   & 41\%/\textbf{40\%}   & 45\%/\textbf{83\%}   & \multicolumn{1}{l|}{59\%/\textbf{82\%}}   \\ \hline

\end{tabular}
\end{table}

\textbf{A closer look.}
We now zoom into the fine\hyp grained categories to drill deeper with regard to the differences between the cascading and control settings. 
We study the difference between the percentage of keyphrases and facts, averaged across hops, as depicted by the circles ($\circ$) in \Figref{fig:tokens_fine}. We find that the differences are largest in what can be considered the most important facts. For example,  the categories \textit{Conclusion\slash General}, \textit{Intervention\slash General,} and \textit{Outcomes\slash General} have the biggest difference for facts labeled as \emph{A} and for keyphrases (recall that there are no conclusion\hyp related keyphrases). The percentage of facts labeled as $C$ and $D$ for these categories is also higher in the cascading scenario. This is interesting as one might expect that the telephone effect should impact peripheral categories such as \textit{Duration} the most, whereas the opposite is the case. Analyzing the telephone effect from this perspective, we find that iterative summarization creates a ``tunnel vision'' effect, where less important information often moves to the fore when multiple summarization steps are involved.

\textbf{The bright side of the telephone effect.} Cascading summaries clearly have the disadvantage of propagating errors. However, they also have a potential advantage: given that one person did an excellent job at summarizing a text, it may be that the text they created is more straightforward for the next person to understand and thus to summarize. To determine the existence and strength of such a hypothetical positive effect, we proceed as follows. For each paper and each hop in the cascading setting, we consider only the summary whose source was the ``best'' of all summaries in the previous hop, where we consider the best summary to be the one with the largest number of $A$-labeled facts. We then compare the percentage of facts and keyphrases per fine\hyp grained category between these summaries and the ones in the control setting as previously done. We show the values for this comparison for each fine\hyp grained category as an upward triangle ($\triangle$) in \Figref{fig:tokens_fine}. Similarly, we also show the values for an analogous comparison where the source summary was the ``worst'' summary, depicted as a downward triangle ($\triangledown$). 

By inspecting facts labeled as \emph{A}, we find that the ``best\hyp ancestor'' scenario preserves facts better or similarly to the control setting (except for the \textit{Conclusion\slash General} and \textit{Intervention\slash General} category). This shows that a summary can, in fact, be a better reference text than the original abstract itself. A potential cause for this improvement might be that, although the telephone setting causes distortions, it also lowers the cognitive load, as summarizing jargon\hyp filled 2000\hyp character abstract is significantly harder than summarizing a 500\hyp character text that was written by someone whose knowledge on the subject may be closer to one's own. On the other hand, the ``worst\hyp ancestor'' scenario highlights the dangers of the telephone effect, where one sloppy summarization may distort the facts present in the text and harm subsequent summaries.

\subsection{RQ2: Hop-to-hop information persistence}
\label{rq2}

Next, we calculate the \textit{conditional chance of survival} for keyphrases and facts across hops. That is, given that a fact or a keyphrase is present in a given hop $k$, what is the chance that it will also be present in hop $k+1$? In the case of keyphrases, for a given summarization process and category, we define the survival probability as the percentage of keyphrases that continue to exist in the summary, out of those that already exist in the reference text. For instance, if there are three keyphrases in the \textit{Outcomes} category in hop $1$, and two remain at hop $2$, the probability of survival for this category from hop 1 to hop 2 is $2/3$.
In the case of facts, we consider the number of statements of a given category that
are fully (fact value $A$) or partially ($B$) preserved
in the source text and continue to be so in the resulting summary.
The mean values along with a linear regression line for different categories can be seen in the first column of \Figref{fig:contintional}. Interestingly, we observe that, while by construction the target length of texts decreases exponentially hop by hop, survival probabilities decrease or increase only linearly.
 
\textbf{Cascading \vs control.} We continue to inspect the first column of  \Figref{fig:contintional}. The probability of survival of keyphrases increases for cascading summaries across hops for all coarse categories. This makes sense intuitively: if a keyphrase was selected for a summary, it is likely that it is relevant and thus will be selected again.
In the control scenario, keyphrases always exist in the reference text (because it is the original abstract), so the probability of survival decreases as target lengths decrease.
Analyzing facts, however, we find a significant difference between our experimental conditions: the probability of survival for facts in the control \textit{and} cascading settings decreases across hops, implying that different dynamics govern distortions associated with lexical \vs\ semantic elements of a text. Whereas survival probabilities increase for keyphrases in cascading summaries, they decrease for facts.
We conjecture that some salient keyphrases have an inherent ``fitness'' for survival, but that this is less true for abstract facts, which may be more or less salient, and thus ``fit'', in their concrete surface forms.

\begin{figure*}[!htb]
    \centering
    \begin{minipage}{.7\textwidth}
        \centering
        \includegraphics[width=\linewidth]{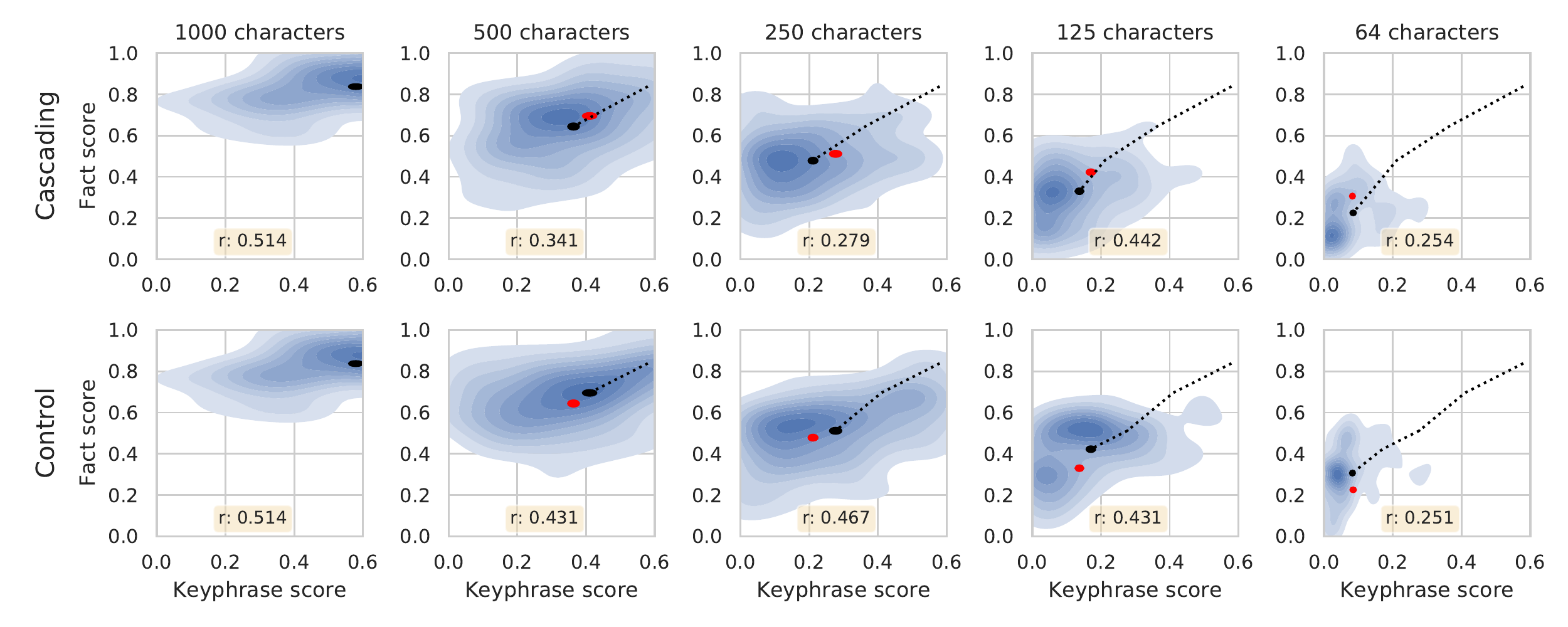}
        \subcaption{}
        \label{fig:summarization_strategies}
    \end{minipage}%
    \begin{minipage}{0.29\textwidth}
        \centering
        \includegraphics[width=\linewidth]{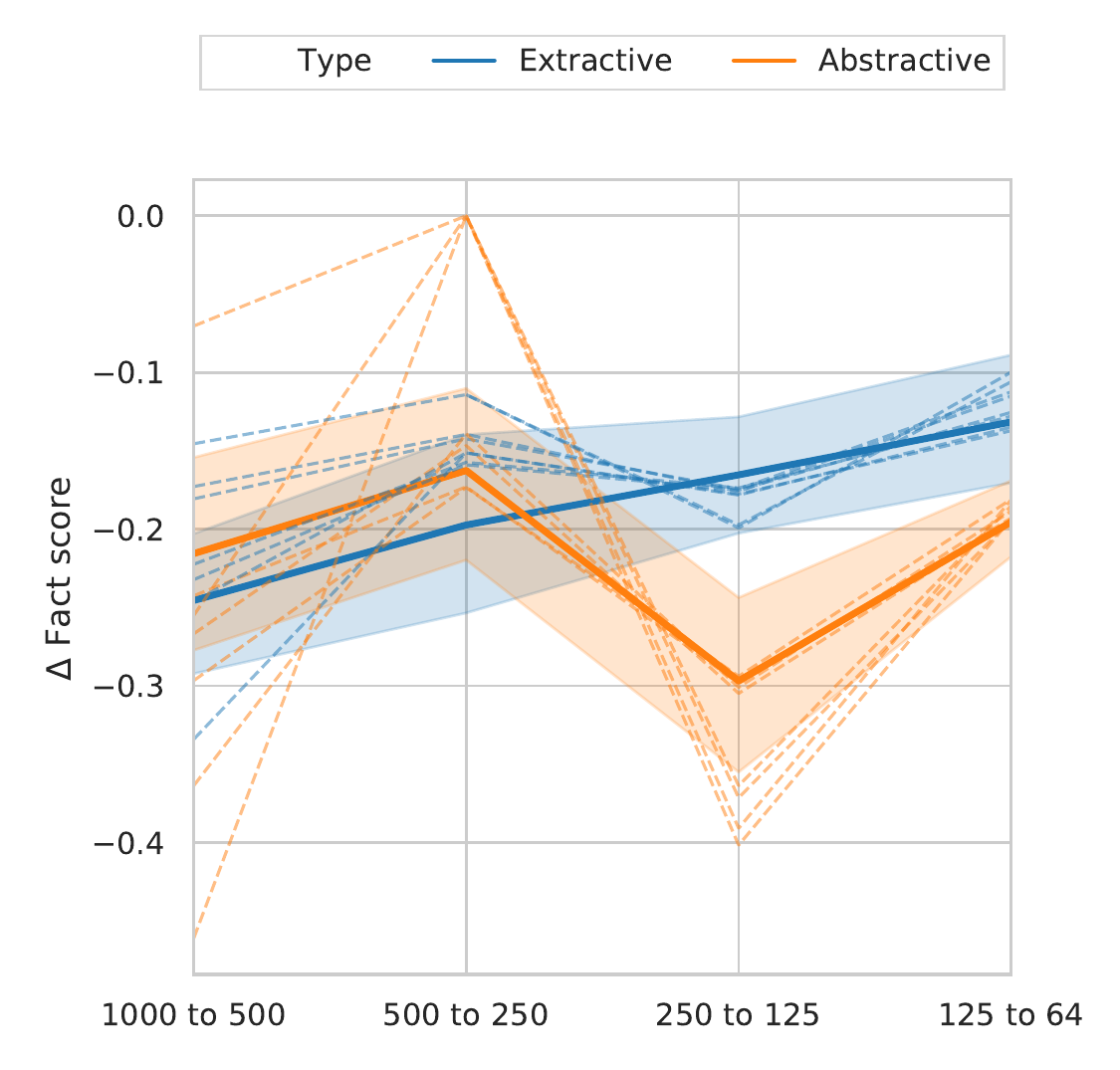}
        \subcaption{}
        \label{fig:summarization_strategies2}
    \end{minipage}
    \caption{(a) Kernel density estimates of summaries according to their fact scores and keyphrase scores for different hops in cascading and control settings. Ellipses mark centroids (height and width: 95\% CIs; black: experimental setting in question; red: other experimental setting).
    (b) Loss of fact score per summarization step for extractive \vs\ abstractive texts in the cascading setting. Lines correspond to multiple combinations of $\alpha \in [0.3,0.7]$ and $\beta \in [0.0,0.15]$ (95\% CIs plotted for $\alpha =0.3, \beta = 0.05$).
    }
\end{figure*}

\textbf{Categories.} Moreover, we can also find differences in the dynamics of distortion across different categories. We observe that categories have different survival probabilities, especially for facts.
For control summaries, the survival chance of \textit{Conclusion} facts is higher than for the facts of other categories. Also, for \textit{Conclusion}, the survival chance in cascading summaries is lower than for control summaries. This suggests that using the original abstract as a reference makes it more likely that the reader will understand the conclusion. Notice that this is different from what we observed in \Secref{rq1}, as there, it could be the case that the conclusion prevailed in the control setting simply because it got lost in the cascade. Here we see that, even when the conclusion fact is present in a given summary in the cascading setting, it is less likely that it will survive.

\textbf{Knowledge questionnaire.} 
We consider the influence of the worker's knowledge questionnaire score (\Secref{sec:dataset_medical}) on the level of distortion. At each hop, we rank the texts according to the score of the workers who summarized them on the questionnaire of the topic associated with the original abstract, and then compare the survival chance of facts and keyphrases of two extremes: the texts summarized by workers whose score is in the first quartile (who performed the best in the test) and the texts summarized by workers whose score is in the fourth quartile (who performed the worst). This is shown in the second column of \Figref{fig:contintional}. We use Chow's test~\citep{dougherty_introduction_2016} to assess whether the coefficients of the regression are significantly different, finding that, in the control setting, workers who scored better in the test lost significantly fewer facts than those who did poorly ($p < 0.01$). 
For cascading summaries, keyphrases survive more for workers that scored well in the text ($p < 0.05$), whereas the difference in the survival of facts is not significant.
A hypothesis for this disparity is that in-depth knowledge on the subject is essential to read the original abstract (as in the control case), but that, once the text has been summarized by someone else (who may not possess that knowledge), the noise is already introduced, and the effect of the level of knowledge fades away. 

\textbf{Readability.} We also consider the influence of the number of difficult words \emph{in the original abstract} on the survival of facts and keyphrases.
We order texts according to the percentage of words that \textit{(i)} have more than one syllable and \textit{(ii)} are not in a list of the most frequent English words (as done for readability metrics such as the Dale-Chall Readability Score~\cite{dale_formula_1948}). We compare the cascades of the four most readable abstracts (first quartile) with the four least readable (fourth quartile). This is depicted in the third column of \Figref{fig:contintional}. We find a different effect to what we observed when inspecting the knowledge questionnaires: facts in more readable abstracts do not a have a significantly different of survival in any setting according to Chow's test. Keyphrases survive more for more readable abstracts in the cascading setting only ($p < 0.05$). Altogether, these results suggest that the telephone effect alters not only the facts and keyphrases of the summaries, but also how other factors influence the distortion processes over these facts and keyphrases.

\subsection{RQ3: Extractive \vs\ abstractive strategies}

Another aspect that may influence distortion effects in information cascades is the summarization strategy employed. Here we compare \textit{abstractive} and \textit{extractive} summarization strategies. To be able to do so, we have to distinguish abstractive and extractive summaries, as well as successful and unsuccessful ones.
To this end, we define \textit{keyphrase scores} and \textit{fact scores} for each summary. The keyphrase score is defined as the percentage of keyphrases that the summary retained from the original abstract, whereas for fact scores we take a weighted average of the percentages of $A$- and $B$-valued facts retained from the original abstract (assigning weight $1$ to $A$-valued facts and weight $0.5$ to $B$-valued facts). With these definitions, we can plot summaries as points in a two\hyp dimensional plane. We depict the kernel density estimate of these points in \Figref{fig:summarization_strategies}.
In this figure, we consider points with higher fact scores ($y$-axis) to be better summaries, and points with higher keyphrase scores ($x$-axis) to be more extractive.
This way, the figure allows us to visually track the quality and extractiveness of summaries.

In each plot of \Figref{fig:summarization_strategies}, we display a black ellipsis marking the centroid of the summary cloud (the height and width of the ellipsis represent 95\% confidence intervals), and a red ellipsis marking the centroid for the other experimental setting.
We also plot the hop-by-hop trajectory of the centroid as a dotted line. Notice the diagonal trajectory: summaries go from better and more extractive to worse and less extractive. Comparing the centroids for the cascading and the control settings, we also see that the centroid of the cascading setting always lies southwest of that of the control setting: cascading summaries are worse as well as less extractive.

In all scenarios, the fact scores are correlated with keyphrase scores (Pearson's $r$ between 0.25 and 0.51),
as can be seen in \Figref{fig:summarization_strategies}. This makes it particularly hard for us to fairly assess summarization strategies, as it may be that the worst summaries are considered ``abstractive'' simply because they do not contain keyphrases (while also not containing the facts). Thus, our analysis has to condition the comparison in a way that we compare summaries that are similar in ``goodness'', but that differ in the level of extractiveness (which we capture using the defined scores).

We proceed as follows: for each hop and abstract in the cascading setting, we match summaries in pairs. Summaries are only matched if \textit{(i)} the relative difference in their keyphrase score is \textit{greater} than $\alpha$ and \textit{(ii)} the relative difference in their fact score is \textit{less} than $\beta$. We greedily match each summary once. Within each pair, we consider the summary with the higher keyphrase score to be more extractive (while of similar quality). We then compare how the fact score of each summary in the pair decreases with the subsequent summarization. The idea is that if extractive texts lose more of their fact score than abstractive texts, this is evidence that extractive summarization is less effective (and vice versa).

We show the results for each of summarization step in \Figref{fig:summarization_strategies2} for several values of $\alpha$ and $\beta$.
The plot suggests that even when we account for the correlation between keyphrases and facts, we observe that extractive summaries are beneficial to latter hops, but similar in the first ones. In our data, thus, although abstracts were often full of jargon and technicalities, trying to change important lexical parts of the text (which we tagged as keyphrases) seems to be detrimental.
Our analysis here is limited as we are considering the survival of keyphrases to be a proxy for extractiveness, which may not necessarily hold in all cases.
\label{sec:results}

\section{Discussion}
\subsection{Summary of findings and implications}

In this paper, we propose an experimental framework for studying message distortion in information cascades and assess the diffusion of information from selected medical abstracts. 
Our analyses suggest that information cascades, as captured by iterative summarizations, distort the examined medical texts lexically and semantically.
Facts and keyphrases frequently are not captured or even contradict the original text. The ``telephone effect'' impacts the most essential information the most, in particular the conclusion of abstracts. Overall, the content of the message after cascading summarization differs considerably from the content after direct summarization.

The telephone effect is, however, nuanced. 
Firstly, it is not necessarily bad: good summaries may serve as stepping stones toward better downstream summaries. Moreover, cascading summaries attenuate the impact of the complexity of the original text as well as the impact of users' topic\hyp specific knowledge.
Our findings suggest that influential platforms or users may have a disparate impact on the quality of information being spread, resonating with the narrative that the rise of online social networks, where messages written by anyone can have far-reaching impact, has diminished the quality of the information~\cite{howell_opinion_2018}.  

We also investigated the success of abstractive \vs\ extractive summarization in information cascades. Even when accounting for the correlation between keyphrases and facts, we still find that extractive summarization performs better. This insight has implications on how scientists and the press should cooperate to convey research to the general public. Our findings suggest that scientific media coverage should make an effort not to distort the key terms with which authors use in their research.

Lastly, although keyphrases and facts are correlated, the conditional probability of survival of keyphrases increases hop by hop in information cascades, whereas it decreases for facts---a finding that should be taken into consideration when modeling the diffusion of information using keyphrases as proxies for associated facts.

\subsection{Related work}

\textbf{Similar methodologies.} Previous work has considered similar experimental settings to study distinct phenomena. Mesoudi and Whiten \cite{mesoudi_alex_multiple_2008} propose a setting similar to our iterative cascades (among others) to study cultural transmission and evolution. Moussa\"id \etal \cite{moussaid_amplification_2015} use the equivalent of what we call a one-hop iterative cascade to study the propagation of risk information.
A similar analysis of how textual content mutates was done in an observational setting by Adamic et al~\cite{adamic_information_2016}.
The researchers studied how memes evolve through Facebook analyzing the adaptation of textual features, as well as their impact on the propagation of the meme.

\textbf{Word of mouth and customer behavior.} Word of mouth is an important phenomenon driving customer behavior. Prior work has studied how the level of customer satisfaction impacts engagement in word-of-mouth information diffusion \citep{anderson_customer_1998} and how effects of word of mouth together with other factors impact customer persuasion \citep{herr_effects_1991}. Word-of-mouth effects have also been analyzed in a networked framework to demonstrate different roles played by weak and strong social ties and the relational properties of homophily on referral behavior \citep{hennig-thurau_electronic_2004}. 

\textbf{\textit{Bona fide} \vs\ intentional information distortion.} Agents involved in word of mouth distort the information even given best intents. In contrast, a rich body of work seeks to detect, model, and prevent intentional message distortion, e.g., the dissemination of misinformation, and fake news \citep{kucharski_post-truth:_2016, tschiatschek_fake_2018, cresci_fake:_2018}. Our experimental setting abstracts away from the complex social media landscape with heterogeneous agents including bots and trolls \citep{varol_online_2017,flores-saviaga_mobilizing_2018}, in order to understand the fundamental patterns that govern information distortion. On the middle ground between best intent and intentional distortion, media outlets distort information in more a more nuanced way \citep{morstatter_identifying_2018}. In \textit{agenda setting,} more attention is allocated to stories that fit a biased narrative, and in \textit{framing,} the facts are more subtly distorted due to how they are presented.

\textbf{Telephone and summarization effects.} The telephone effect and the summarization effect have been studied from a linguistic standpoint.
Breck and Cardie~\citep{breck_playing_2004} note that facts, events, and opinions appearing in news articles are often known only second- or third-hand. Agents reporting them resort to using two kinds of expression that can filter information: perspective and speech expressions. They propose a learning approach that correctly determines the hierarchical structure of such information filtering expressions emerging due to the telephone effect. 
More recently, Gligori\'c et al.~\cite{gligoric_how_2018} found that length constraints cause Twitter users to summarize their content. This process significantly alters linguistic aspects of the text, such as the use of abbreviations, contracted forms, and articles.

\textbf{Science communication.} Media coverage of science has been investigated by communication research. The focus of the field is on scientists' attitude towards the media and on patterns of interactions with journalists and other key players such as news organizations and science information professionals \citep{weigold_communicating_2001, peters_gap_2013}. Evidence suggests that most scientists consider visibility in the media important and responding to journalists a professional attitude that is reinforced by universities and other science organizations. Exaggeration in the news is strongly associated with exaggeration in press releases, and improving the accuracy of academic press releases could represent a key opportunity for reducing misleading health-related news \citep{sumner_association_2014}.

\textbf{Summarization and paraphrasing.} Finally, the summarization task at the heart of our experimental design is tightly related to the classic natural language processing tasks of paraphrasing \cite{yin_convolutional_2015, hu_convolutional_2014, resnik_improving_2010} and summarization \cite{mani_summarization_2009,nallapati_abstractive_2016}. Prior research introduced the distinction between abstractive and extractive summarization approaches upon which we rely in some of our analyses \citep{hsu_unified_2018}.


\subsection{Limitations and future work}
Our experimental setup allows us to finely measure and characterize message distortion effects, which would be difficult in observational setups. Still, it remains unclear to what extent our findings generalize to a wider spectrum of real-world information cascades, where three important distinctions are likely to have an impact: 
\textit{(i)} the sequence-of-laypeople mechanism used here may differ from the way information diffuses in the real world; 
\textit{(ii)} different kinds of information (\eg, news) may be subject to different dynamics; 
\textit{(iii)} the assumed \textit{bona fide} scenario may not hold in all cases.

Future work will face the challenge of tracking message distortion in real-world cascades (such as the one described in \Secref{sec:intro}). By taking advantage of methods and insights developed in this paper, we plan to establish the extent to which our findings reflect the complex universe of social media, thus removing the \textit{bona fide} assumption. Moreover, we will explore whether the observed effects can be modeled mathematically to predict which lexical and semantic units of a text that are likely to be distorted.



\label{sec:discussion}


\section*{Acknowledgements}
We gratefully acknowledge support from CNPq, Atmosphere, a Google Research Award for Latin America (Manoel Horta Ribeiro), and a Google Faculty Research Award (Robert West).
We would also like to thank Lucy Li for helpful discussions.

\bibliographystyle{ACM-Reference-Format}
\bibliography{Zotero} 

 \end{document}